\documentstyle[12pt,aasms4]{article}

\lefthead{Henriksen}
\righthead{X-ray Constraints on Cluster Magnetic Fields}

\begin{document}

\title{X-ray Constraints on Cluster Magnetic Fields}

\author{Mark Henriksen}
\affil{Physics Department, University of North Dakota,
    Grand Forks, ND 58202-7129}
\centerline{mahenrik@plains.NoDak.edu}

\begin{abstract}
Lower limits to the intracluster 
magnetic field are derived
from modeling broad band X-ray spectra of 6 galaxy clusters with radio
halos (A401, A754, A1367, Coma (A1656),
A2256, and A2319). The 0.7 - 10 keV
spectra for all clusters were obtained
with the Advanced
Satellite for Cosmology and Astrophysics ({\it ASCA})
Gas Imaging Spectrometer (GIS). For A401, A754, A1367, and A1656,
data from the {\it HEAO1-A2} High Energy Detector (HED) (2-60 keV) 
was included in a joint fit to
further constrain the allowed non-thermal component.
Upperlimits to the non-thermal X-ray emission are
a factor of $\sim$10 improvement over previous values.  
Lower limits to the average magnetic field derived
from these upperlimits and the radio spectral parameters
range from 0.08 - 0.30$\mu$Gauss. A seed intracluster
field amplified by turbulence from galaxy motion
through the intracluster medium should have an average value in
the range
$\sim$0.1-0.2 (Goldman \& Rephaeli, 1991; De Young, 1992),
thus 
the lack of any detections makes it unlikely that field amplification
is due to galaxy wakes. In addition, all of these clusters have 
an asymmetric surface brightness
distribution while several also
have X-ray spectroscopic evidence of a merger. We suggest
that cluster mergers rather than galaxy wakes are the
source of the magnetic field amplification. 

\end{abstract}

\section
{Introduction}

The presence of a magnetic field and a population of relativistic electrons
is inferred from modeling the radio halo in the Coma cluster (A1656)
which is well described
by a synchrotron spectrum over the approximate range, 10MHz - 3GHz 
(Giovanni et al. 1993). Other clusters of galaxies are also
known to have diffuse radio halos: A401, A754, A1367, 
A2255, A2256, A2319 (references
in Table 5). Non-thermal 
X-ray emission is expected from inverse Compton scattering
of the cosmic microwave background photons off of the relativistic 
electrons. Detection of the predicted non-thermal component
is complicated since X-ray observations of galaxy 
clusters are dominated by a hot intergalactic medium
with continuum emission well described in the X-ray regime 
by thermal bremsstrahlung. 
Previous investigations have focused on 
energies above the exponential decay of the thermal spectrum,
producing upper limits on non-thermal X-ray emission from
some of the clusters known to have diffuse
radio halos: A401, A2255, and A2256 (Rephaeli \& Gruber 1988),
A1367, A2319 (Rephaeli, Gruber, \& Rothschild 1987) using HEAO1-A4
observations, and most recently for the Coma cluster, using the OSSE experiment 
on the Compton Gamma-ray
Observatory (Rephaeli, Ulmer,
\& Gruber 1994).
The upper limit on
non-thermal X-ray emission together with the radio 
spectral parameters are used to place a lower limit on the average 
magnetic field, $<$B$>$. These papers have reported lower limits to
$<$B$>$ for the clusters as high as $\le$0.15 $\mu$G.

The origin and amplification mechanism 
of cluster magnetic fields is somewhat of
a mystery. A primordial
seed field amplified by gravitational collapse of
a proto-cluster results in too weak of a field to give
the observed Faraday Rotation measures (RM)
(Ruzmaikin, Sokoloff, \& Shukurov 1989).
Evidence of
metals in the intracluster medium out to z$\ge$0.3 (Mushotzky
and Loewenstein 1997) suggests that
a seed field would originate
in ejected galactic material.
Turbulence from the motion of galaxies in the intracluster
medium would amplify the
seed field to the observed RM levels (Ruzmaikin, Sokoloff,
\& Shukurov 1989). However, this mechanism
can only produce an average field of $\sim$0.1-0.2
$\mu$Gauss (Goldman \& Rephaeli 1991; De Young 1992). If
cluster magnetic fields are typically higher than $\sim$0.1-0.2,
then some other amplification mechanism would be needed.
Previous upperlimits to $<$B$>$ have been within this
range or below and are thus inconclusive.
Rotation Measures, though indicating a magnetic
field in excess of this value, may
be biased by local density peaks in a clumpy intracluster
medium and may not give an accurate measure of the cluster
magnetic field. Combined X-ray and 
radio measurements, such as we present in this
paper, give a more robust measurement of the
average cluster magnetic field.

If strong magnetic fields are present in galaxy
clusters, they are expected to impact a number
of other research areas, such as the relationship of 
gas mixing (Valinia et al. 1996) during
cluster evolution
to the distribution of metals
in the intracluster medium as well as cluster mass measurements
derived from the radial gas pressure distribution
(Makino 1997). In particular,
 it has been suggested that
the presence of 
an equipartition magnetic field in the A2218 cluster
might reconcile the apparent discrepancy between the total cluster
mass inferred from gravitational lensing and that inferred from the X-ray
emitting intracluster medium (Loeb \& Mao 1994; Ensslin et al. 1997).  

In this paper, we present new upper limits on the non-thermal X-ray
component in six clusters of galaxies with diffuse radio halos 
using {\it ASCA} GIS and {\it HEAO1-A2} HED spectra (Section 2). In Section 3, 
the procedure is given for calculation of the
average magnetic field from the X-ray and radio parameters. 
In Section 4, we discuss the impact of the average magnetic field
lower limits on theories of magnetic field amplification.
We assume a value H$_{0}$ =
50 km sec$^{-1}$ Mpc$^{-1}$ in all of the
calculations.  

\section
{X-ray Spectral Observations}

The significant limits these data place
on the non-thermal X-ray component
are a result of 
utilizing observations made with the {\it ASCA}
(Tanaka, Inoue, \& Holt 1994) GIS
(Ohashi et al. 1996; Makishima et
al. 1996) and to the technique
of fitting multiple data sets to expand the energy band.
In a joint fit, the {\it ASCA} and {\it HEAO1-A2} 
 data provide broad band
spectral coverage $\sim$(0.7 - 60 keV). The GIS tightly
constrains the gas temperature and also limits the
allowable contribution of the power law component to the
0.7 - 10 keV spectrum. The HED data provides modest
sensitivity in the hard X-ray regime, extending the
spectrum out to 50 - 60 keV.  The HED exposure times and count rates
are given in Table 1.
The HED has a field of view of
3$^{o}$x3$^{o}$ and an energy resolution of $\sim$20\%
at 6 keV. The HED has an effective area of 429 cm$^{2}$
and very accurate background subtraction utilizing
offset pointings. 
The cluster with the highest signal-to-noise 
data, the Coma cluster, has $>$ 2\% random statistical
error in the channel with the highest signal-to-noise.
The systematic error in the energy calibration
is $<$ 1\% and is not included in the spectral modeling
because it is negligible compared to the statistical error.
The characteristics of the {\it HEAO-1 A2}
data are more fully described in Henriksen \& White (1996)
and references therein. 

The GIS2 event files, undergone filtering using REV2 standards,
were used with the latest (v4.0) response matrices. The v4.0
response matrices provide improved modeling of the low energy
pulse height distribution. The high energy response calibration
is dominated by statistical errors for sources of several
counts sec$^{-1}$ or less (GIS News No. 2, 1995). Table 1
shows that the GIS count rates for the clusters,
which range from 0.8 - 5.25  counts sec$^{-1}$, are
relatively low and should be dominated by statistical
errors at high energy.
Comparison of the spectral parameters measured 
independently by the HED and GIS show good agreement (see Table 2)
also indicating that it is unlikely that systematic errors in the
GIS calibration dominate at high energies.

Blank sky GIS images were used for background subtraction since the
cluster emission may cover the entire detector, thus precluding
use of a source free region of the detector for background. The same image
region on the GIS detector is used to extract the background
region as is used to extract the source spectrum. Spectra were modeled
using XSPEC version 9.0 (Arnaud 1996).
 The model used to fit the data consists of a Raymond
and Smith plasma code (1977) with
absorption cross sections (Morrison \& McCammon 1983),
solar abundance ratios (Anders \& Grevesse 1989), and a power law
component.
The normalization on the thermal component
is proportional to the emission integral which will vary with the
size of the region modeled. 
For this reason, the HED and GIS 
	normalization on the thermal component are not tied. The derived
	90\% confidence range on the thermal normalizations are given in Table
	3 for the joint fit. They are generally in agreement except in the
	case of A1367. A1367 has a very irregular morphology with a 
	broad surface brightness distribution compared to the other clusters.
	It is clearly bimodal. It is also, along with Coma, the closest. Thus, the thermal normalization, 
	 which is proportional to  the
	emission integral,
	is quite different in each data set because of the diffuse emission
	outside of the GIS FOV. Other free parameters are the temperature, 
abundance, and
column density. The abundance, temperature, and column density
are tied, meaning the HED and GIS data are required to have the
same value of these parameters. The power law
photon spectral index ($\alpha_r$ + 1)
is fixed at the value calculated from the radio energy spectral
index
($\alpha_r$)(see Table 5). Only data above 0.7 keV is used and
the data is grouped to contain at least twenty counts in each bin to
increase the reliability of the results obtained from a $\chi^{2}$ fit
(Nousek \& Shue 1989). 
The GIS spectrum is obtained from a
region of radius 18 arcmin, which is chosen to enclose the  reported radio
halo, radius of 5 - 15 arc min (see Table 5), for every cluster. 
Since both data sets
contain the non-thermal component, we tied these two
normalizations. This is justified because, as we show below,
the absolute photometric calibration of two spectra
differ by less than 6-7\%.

Our method for constraining the NT component
differs from earlier studies in the following.
Earlier studies have fixed the thermal component 
at the emission
weighted value and looked for residuals at higher energy. 
Our method provides better modeling of 
the thermal component by allowing it to vary
rather
than be fixed at a single value. This also
makes it possible to use {\it ASCA} data above 0.7 keV, 
the highest signal-to-noise spectral
data available for galaxy clusters, to help
constrain both the thermal and power law component. 
Also,
if multiple temperature components are
present, then the GIS, with greater sensitivity at higher
energy than other available data (e.g., 
the {\it ASCA} SIS or {\it ROSAT} PSPC) will most accurately model the
hottest gas component and therefore
provide the most accurate separation
of the thermal and non-thermal (NT) components.
The 90\% confidence upper limit 
is measured by stepping the normalization
of the NT component through a range of
values while minimizing chi-square ($\chi^{2}$). 

To assess the validity of our approach to modeling the
non-thermal emission, we performed several consistency
checks regarding the calibration of the {\it HEAO1-A2} HED and {\it ASCA}
GIS detectors.
To compare the calibration of the data,
a Raymond and Smith model was fit both independently 
and simultaneously to both data sets. The independent fit
tests whether or not the spectra are the same shape while the simultaneous
fit, with a subsequent calculation of the flux 
measured independently by each detector, tests
the absolute photometric cross-calibration of the two data sets. 
Table 2 gives the results of independent fits, including the 90\% limits
on the cluster column density, temperature, and abundance.
Comparison of the GIS
and HED measurements show that the fit parameters are consistent.
The best test of consistency is the Coma cluster where the HED
signal-to-noise is highest and the parameters are well constrained
by each data set. The parameters measured by the GIS and HED
for Coma are in excellent agreement. To evaluate the relative photometric
calibration of the two data sets, a single thermal component was
fit to the A401 data sets simultaneously and the normalization
of each data set allowed to vary (as was done when a powerlaw
was also included). 
A401 was chosen since it is the most distant, z=0.0748,
and the nominal extent of cluster X-ray emission, $\sim$3 Mpc
is within the GIS field-of-view.  
The flux in the 2-10 keV band, the overlapping
energy of the two data sets, is: 1.10$\times$10$^{-2}$ photons cm$^{-2}$
sec$^{-1}$ in the GIS and 1.03$\times$10$^{-2}$ photons cm$^{-2}$
sec$^{-1}$ in the HED. The flux scales as the normalizations and
90\% confidence limits on the normalizations overlap.
Thus, the HED and GIS, for our purposes, are well calibrated, to $\sim$6-7\%,
and in excellent agreement.

Table 2 gives the results of the joint fit.
No NT emission is required to 
obtain a good fit. Figures 1 - 6 show $\chi^{2}$ vs. energy for the
best fit thermal spectrum with the upperlimit
to the NT component added. Also shown are the model components,
both individually and separate. In general, the entire thermal
spectrum is distorted when a powerlaw is added.
For a low temperature cluster such as A1367, the powerlaw
becomes dominant in the spectrum above 20 keV. The 0.7 - 2 keV 
part of the spectrum also constrains the powerlaw. 

The possibility that the effect of
additional absorption beyond
that accounted for by  galactic neutral hydrogen affects the fit
was also investigated. 
Table 3 contains the 90\% confidence
range on column density measured by the GIS as well as the Galactic
n$_{H}$ column density (Stark 1992). For all of the clusters,
the agreement is excellent. This is also consistent with White et al.
(1991) who found, using the Einstein Solid State Spectrometer, 
that of these 6 clusters, 
only A401 may show any evidence of additional absorption. These clusters
also show substructure and are likely merging systems. Mergers 
anti-correlate with the presence of a cooling flow (which
strongly correlate with absorption intrinsic to the cluster).

No NT components were detected, consistent with past 
analysis, however, a factor of $\sim$10 more sensitive lower limits to the
average cluster field were derived
for 5 clusters: A401, A1367, A1656, A2256, and A2319.
For A754, this is the first measured NT upperlimit.
The upper limits to the NT X-ray flux are given in Table 4 along with the
column density, temperature, and abundance. The monochromatic
X-ray flux is calculated in photons cm$^{-2}$ sec$^{-1}$ keV$^{-1}$
from the NT normalizations given in Table 2 and compared to previous
upperlimits in Section 4. All values
are reported with 90\% confidence.

\section{Magnetic Field Calculation}

In this section, we begin with basic equations which describe
the radio and NT X-ray emission mechanisms and show how they
are combined to constrain the average magnetic field
using observables. The radio halo spectra 
are described by a powerlaw function of the
form,
\begin{equation}
S(\nu) = A\nu^{-\alpha_{r}},
\end{equation}
Where S is the radio flux and $\alpha_{r}$ is the energy spectral index.
Values of A are given in Table 5.
The synchrotron power per volume per frequency (P$_{s}$) is given by equation 6.36 in
Rybicki and Lightman (1979) which is rewritten here in terms of
frequency ($\nu$) as,
\begin{equation}
P_{s}(\nu) = K_{1}C<B_{P}>^{\frac{p + 1}{2}}\nu^{\frac{1-p}{2}},
\end{equation}
where
\begin{equation}
K_{1} = \frac{\sqrt{3}q^{3}}{m c^{2}(p + 1)}\Gamma(\frac{p}
{4} + \frac{19}{12})\Gamma(\frac{p}{4} - \frac{1}{12})[\frac{2\pi mc}{3q}]^{-\frac{(p-1)}{2}}.
\end{equation}
The relativistic electron distribution,
N($\gamma$) = C$\gamma^{-p}$, specifies C where p is the electron
spectral index and $\gamma$ is Lorentz factor. P is calculated from 
$\alpha_r$, p = 2$\alpha_r$ + 1. B$_{P}$ is the magnetic field component which is
perpendicular to the line of sight.
In equation (3), q is the electron charge, c is the speed of light,
m is the rest mass of the electron, and $\Gamma$ is the Gamma function.

The expression for the
inverse Compton
emission  ($P_{c}$) from
cosmic microwave background radiation (CMBR)
scattering off of the relativistic electrons is
given by equation 7.31 in Rybicki and Lightman (1979),
\begin{equation}
P_{c}(\epsilon) = K_{2}C(kT)^{\frac{p + 5}{2}} \epsilon^{\frac{1-p}{2}},
\end{equation}
where T is the temperature of the CMBR, $\epsilon$ is the photon
energy, and
\begin{equation}
K_{2} = \frac{8\pi^{2}r_{0}^{2}D(p)}{h^{3}c^{2}}.
\end{equation}
In this equation, r$_{0}$ is the classical electron radius and h is Planck's constant,
\begin{equation}
D(p) = J(p)\Gamma[\frac{p+5}{2}]\psi[\frac{p+5}{2}],
\end{equation}
\begin{equation}
\psi(s) = \sum_{n=1}^{\infty} n^{-s},
\end{equation}
\begin{equation}
J(p) = 2^{p+3}[\frac{p^{2} + 4p + 11}{(p + 3)^{2}(p + 5)(p + 1)}].
\end{equation}
An expression for the NT X-ray flux (F$_{c}$)is derived in terms of observable
quantities from equation (4) by eliminating C using
equation (2). P$_{s}$ and P$_{c}$ are integrated over volume, and divided
by 4$\pi$D$^{2}$, where D is the distance to the cluster.
\begin{equation}
F_{c} = \frac{K_{2}}{K_{1}}(kT)^{\frac{p+5}{2}}<B_p>^{-\frac{p+1}{2}}
A\smallint\epsilon^{-\frac{p-1}{2}}d\epsilon.
\end{equation}
The energy band of the observed flux determines the integration
limits. Here, the limits were chosen to be 2-10 keV when the {\it ASCA} data is used
alone (A2256, A2319) and 20-60 keV when {\it HEAO1-A2} data is also fit. The
flux is given in Table 4 in these energy bands.
The spectral fit utilizes the full range of the data sets,
0.7 keV - 60 keV, so that 2-10 keV and 20-60 keV are nominal energy
ranges for calculation of the hard X-ray fluxes. These energy
bands are chosen to facilitate comparison with earlier
measurements such as those referenced in Section 1.
The temperature of the CMBR is taken
as 2.7 K. This calculation is independent 
of the specific cosmic-ray
electron distribution, size of the emitting region, or distance
to the cluster. Using the radio data alone to determine
$<B_p>$, (i.e., equation
(2)) would additionally require constraining C explicitly as well as
the distance to the cluster and the size of the emitting
region.
The calculated $<B_p>$ is given in Table 4 and ranges
from 0.08 (A401) - 0.30 (A2256)$\mu$Gauss. 

\section{The Amplification of Cluster Magnetic Fields}

Using archival data, we have failed
to detect non-thermal X-ray emission in the broad 0.7 - 60 keV
band formed by {\it ASCA} and {\it HEAO-1 A2} observations
of clusters with radio halos. The upperlimits on non-thermal
X-ray emission derived from fitting these data were used with
the radio parameters to obtain lower limits on the
average magnetic field. 

The method presented, 
modeling non-thermal emission as 
inverse Compton scattering of CMBR photons off of relativistic
electrons,
has been used to successfully detect
high energy emission in sources other
than clusters with radio halos: for
example, 
the radio lobes of Fornax A (Kaneda et al. 1995; Feigelson
et al. 1995) and possibly several galaxy clusters.
Fornax A, located in the outer regions of a galaxy cluster,
has a value of 2-3$\mu$Gauss
for the magnetic field associated with the radio lobes. 
Sarazin and Lieu (1998) suggest that the emission
detected from several galaxy clusters using 
the Extreme Ultraviolet Explorer ({\it EUVE})
(Lieu et al. 1996)
may be also be non-thermal emission. The relativistic 
electrons which give the {\it EUVE}
emission in galaxy clusters are of lower energy than those 
that produce the 
radio halos and their radio emission should be unobservable
from Earth (Sarazin and Lieu 1998) as a radio halo. The
{\it EUVE} emission may involve the low energy tail of the
same population of relativistic electrons which produce
the radio halos providing important information on the spatial
distribution of the 
magnetic field and population of relativistic electrons. 

Lower limits to the average cluster magnetic field values 
reported here are 
higher than
earlier measurements for most of the clusters (A754 was not measured
earlier). Though the HEAO1-A4 data used
in these studies has the advantage of covering a higher energy
band, 13 - 125 keV, the detectors have an effective area 
5 - 10 times lower than the {\it HEAO1-A2} HED. They also
did not model the thermal spectrum. The {\it ASCA}
data not only models the thermal spectrum well,
it provides a tight constraint on the overall shape
of the spectrum, thus providing important limits on the
flux of any additional NT component.
For comparison, the 
upper limits on the NT flux at 30 keV are given for the 5
clusters in common in units of 10$^{-6}$ photons cm$^{-2}$ sec$^{-1}$ 
 keV$^{-1}$ as (A2,A4): A401 (2.2,19.0), 
A1367 (1.8,21.0), A1656 (1.9,
24.0), A2256 (0.37, 5.), and A2319 (2.1, 22.). 
For comparison, the 
lower limits to the average magnetic field are given for the 5
clusters in common as (A2,A4): A401 (0.08,0.04), 
A1367 (0.19,0.056), A1656 (0.26,
0.11), A2256 (0.30, 0.15), and A2319 (0.23, 0.11). 
Our results indicate
that average cluster fields are typically at least 0.2 $\mu$Gauss
and several are in excess of this value.

These values for the average magnetic field in the intracluster
medium begin to constrain theories for their origin.
It has been hypothesized that a seed field is supplied through
gas loss by galaxies and is then amplified by the turbulent
motion of
galaxy wakes. Estimates of the mean field 
achievable by this amplification process
range from $\sim$2$\mu$G (Ruzmaikan
et al. 1989) to 0.1 - 0.2 $\mu$G (Goldman and Rephaeli 1991).
Amplification of seed fields has also been considered
by De Young (1992) using time-dependent evolution of magneto-hydrodynamic
turbulence from galaxy wakes. He found that it is very
difficult to produce $\mu$Gauss fields with this mechanism
and that radio halos 
must result from extreme
conditions in this scenario.
Goldman and Rephaeli (1991) use a lower, more realistic 
value for the efficiency of
conversion of kinetic energy into magnetic field
energy thus getting ambient fields in the range of 0.1 - 0.2.
Their work suggested that the magnetic fields 
amplified by galaxy wakes should be confined to this
range and should have been detectable.
Our lack of detections begin to question galaxy wakes as
a source of the magnetic field energy since they cannot
amplify the magnetic fields to the observed lower limits. Larger fields require 
a greater source of kinetic energy. 

Part of the legacy of the {\it ROSAT} observatory
is the discovery that X-ray
substructure in galaxy clusters is prevalent. 
All of the
clusters have asymmetric X-ray
morphology similar to that seen in hydrodynamical simulations
of cluster mergers (Roettiger, Loken, and Burns 1997). At least
three of the clusters, A754 (Henriksen and Markevitch 1996), 
Coma (Honda et al. 1996), and A1367 (Donnelly et al. 1998)
have spectral evidence of a merger as well.
Some of the 
other clusters 
also show additional evidence of a merger: 
A401 (Fujita et al. 1996; Fabian, Peres, and
White 1997), 
A2256 (Miyaji et al. 1993; Briel and Henry 1994; Roettiger,
Burns, and Pinkney 1995), and A2319 (Markevitch 1996). 
The similarity between the direction of the 
elongation in the radio and X-ray
maps of the Coma cluster also suggests a connection between
the merger and the radio halo (Deiss et al. 1997).
Burns et al. (1994)
found that the energy from the merger of the NGC 4389
group is sufficient to power the radio halo. 
Bohringer et al. (1992) also found that the
available energy specifically from the accretion of groups
near the core may be enough to reaccelerate cosmic
ray electrons and amplify the magnetic field to maintain
the halo. 
The energy associated with subcluster merger
will amplify the magnetic field and accelerate
the relativistic electrons to produce the observed
level of synchrotron emission. Okoye
and Onuora (1996) found that only a small amount
of boosting is necessary to  explain the observed
radio luminosity of the Coma C source (radio halo).
Similar conclusions have been found for the
A2319 cluster (Feretti, Giovanni, Bohringer 1997).
It is also important to note that
amplification by subcluster merger is fundamentally
different than from galaxy wakes since galaxy motions
are continual and subcluster mergers are isolated in
time. That the merger will have a lifetime much less
than the cluster lifetime means that if the merger
accelerates the cosmic-ray electrons, then the radio
halo will also have a lifetime less than the cluster.
An 
explanation for merging clusters with no radio
halo may be related to the large amount kinetic energy
available in a major merger. The turbulent motions
in major mergers may amplify the seed field
to a strong enough level so that the synchrotron
cooling time is substantially
less than the merger lifetime. In this scenario, we would
only see those radio halos from select mergers: those
that are strong enough to make radio halos yet 
weak enough so that synchrotron lifetime is long
enough for the radio halo to be observable.

\section{Conclusions and Future Prospects}

Calculation of the average magnetic field in galaxy clusters
with radio halos, as outlined in this
paper, is straight-forward  and depends on observable radio and X-ray 
parameters. Lower limits to the average
magnetic field range from 0.08 - 0.30 $\mu$G. They are higher than previous
measurements due to our fitting a combination of the 
{\it ASCA} GIS
and {\it HEAO1-A2} HED which make a broad band, 0.7 - 60 keV spectrum.
The highest values, $\sim$0.30 $\mu$G, are higher than the
range, 0.1 - 0.2 $\mu$G, believed to be amplified
by galaxy wakes; several others are in the range of 0.1 - 0.2 $\mu$G,
yet are undetected. If future observations of these clusters
show that typical average
cluster magnetic fields are $>$ 0.2 $\mu$G then amplification
by subcluster merger rather than galaxy wakes may be more
plausible. The higher magnetic field created by a major merger
also results in a shorter
synchrotron cooling time for the cosmic-ray electrons and therefore
a shorter lifetime for the radio halo. This would
account for the fact that radio halos are rare
while clusters with substructure are common.

The Hard X-ray
Detector (HXD) (Takahashi et al. 1996) to be flown on ASTRO-E will 
have $\sim$10 times the effective
area at 10 keV compared to the {\it HEAO1-A2} and extend to higher energy.
This will 
allow detection or much better limits on cluster magnetic fields to be set.
These new limits to the cluster magnetic field will be able to determine the strength
of cluster magnetic fields and severely constrain theories
for their origin.

\acknowledgements

I am thankful to the referee for suggestions on improving this manuscript
and to Dan Harris and William Schwalm for their suggestions as well.
I acknowledge generous support from the National Science Foundation Grant
AST-9624716.

\clearpage

\begin{deluxetable}{ccccc}
\footnotesize
\tablewidth{0pt}
\tablecaption{X-ray Observations}
\tablehead{
\colhead{Cluster} & \colhead{HED Seconds} & \colhead{HED Counts sec$^{-1}$} &
\colhead{GIS Seconds} & \colhead{GIS Counts sec$^{-1}$}
}
\startdata
A401 & 3990 & 2.32$\pm$0.11 & 32322& 1.39$\pm$0.0076\nl
A754	&4957&1.43$\pm$0.050& 19716&1.82$\pm$0.010\nl
A1367 &4520&2.86$\pm$0.056 & 10646&0.84$\pm$0.0094 \nl
Coma &4247& 8.34$\pm$0.072&9440 & 5.25$\pm$0.024\nl
A2256 &-&-	&  34726 &  1.57$\pm$0.007  \nl
A2319 &-&   - &  14336&2.41$\pm$0.013   \nl
\enddata
\end{deluxetable}
\clearpage
\begin{deluxetable}{ccccccc}
\footnotesize
\tablewidth{0pt}
\tablecaption{Comparison of Best Fit Parameters}
\tablehead{
\colhead{Cluster} & \colhead{Data Set} & \colhead{n$_{H}\times10^{22}$cm$^{-2}$} & \colhead{kT(keV)} &
\colhead{Abundance} & NT Normalization\tablenotemark{a}  & $\chi^{2}$/dof 
}
\startdata
A401 	&  GIS & 0.10 - 0.14	& 7.67 - 8.71	& 0.21 - 0.31 & - 	& 563/609 \nl
	&  HED & 0.0 - 2.76	& 5.59 - 10.90 	& 0.0 - 0.64  &	-	& 53.9/42	\nl
	&  Joint &0.10 - 0.14	& 7.68 - 8.68	 &	0.20 - 0.32 & 	$<$ 0.0077&	617/655 \nl 
A754	&  GIS &0.042 - 0.084 & 9.50 - 11.22	 &0.16 - 0.30&  - & 539/528 \nl
	& HED &  0.0 - 0.51	& 7.29 - 13.89	& 0.0 - 0.60 & - & 12.5/11\nl
	&  Joint &0.041 - 0.081 & 9.67 - 11.28 & 0.15 - 0.29 & $<$ 0.0024 & 552.1/542 \nl
A1367 	& GIS & 0.0 - 0.023 	& 3.3 - 3.7	& 0.17 - 0.40 & - & 306.4/268\nl
	&  HED & 0.0 - 8.94 & 3.17 - 3.83	& 0.10 - 0.23 & - & 66.5/47\nl
	& Joint & 0.0 - 0.022 & 3.4 - 3.7	& 0.14 - 0.25 & $<$0.00862 & 374.9/318\nl	
Coma & GIS & 0.0-0.015&	8.69- 9.49 	& 0.22 - 0.33 & - & 543.3/564\nl
	& HED & 0.00 - 0.78 & 7.92 - 8.70	& 0.14 - 0.27 & - 	& 50/52	\nl
	& Joint & 0.0 - 0.024 & 8.42 - 9.03 	& 0.21 - 0.29 & $<$0.0055	& 600/619 \nl
A2256	& GIS & 0.029 - 0.063 & 6.74 - 7.39 	& 0.16 - 0.25	& $<$0.005 & 682.9/605 \nl
A2319	& GIS & 0.065 - 0.11 	& 9.18 - 10.67	& 0.18 - 0.32	& $<$0.0075 & 562.2/534 \nl
\enddata
\tablenotetext{a}{Non-thermal upperlimit in photons cm$^{-2}$s$^{-1}$keV$^{-1}$ at 1 keV}
\end{deluxetable}
\clearpage

\begin{deluxetable}{ccccc}
\footnotesize
\tablewidth{0pt}
\tablecaption{Data Comparisons}
\tablehead{
\colhead{Cluster} & \colhead{GIS N$_{H}$} & \colhead{Galactic N$_{H}$} &
\colhead{GIS Normalization}\tablenotemark{a} & \colhead{HED 
Normalization}\tablenotemark{a}
}
\startdata
A401 & 0.10 - 0.14 & 0.105 & 0.080 - 0.083 & 0.067 - 0.078 \nl
A754	& 0.042 - 0.070 & 0.045 & 0.099 - 0.10 & 0.055 - 0.062 \nl
A1367 & 0.0 - 0.023 & 0.021 & 0.042 - 0.051& 0.52 - 0.77 \nl
Coma & 0.0- 0.015 & 0.009 & 0.28 - 0.29 & 0.26 - 0.28 \nl
A2256 & 0.029 - 0.063 & 0.042 & - & - \nl
A2319 &0.065 - 0.11 &  0.083 & - & - \nl
\enddata
\tablenotetext{a}{Normalization of thermal component}
\end{deluxetable}
\clearpage

\begin{deluxetable}{ccc}
\footnotesize
\tablewidth{0pt}
\tablecaption{X-ray Spectral Constraints}
\tablehead{
\colhead{Cluster}           & 
\colhead{Compton Flux (10$^{-12}$ ergs cm$^{-2}$ sec$^{-1}$)} & \colhead{Average Magnetic Field($\mu$Gauss)} 
}
\startdata
A401 	& 	 $<$3.3\tablenotemark{a} & $>$0.08 \nl
A754	&	$<$1.5\tablenotemark{a} & $>$0.29	\nl
A1367 & 	$<$2.6\tablenotemark{a} & $>$0.19\nl
Coma &	$<$2.9\tablenotemark{a}& $>$0.26\nl
A2256 & 	$<$4.1\tablenotemark{b}& $>$0.30 \nl
A2319	& 	$<$10.6\tablenotemark{b}& $>$0.23 \nl
\enddata
\tablenotetext{a}{HEAO1-A2 and ASCA data, 20-60 keV}
\tablenotetext{b}{ASCA data alone, 2-10 keV}
\end{deluxetable}
\clearpage

\begin{deluxetable}{ccccc}
\footnotesize
\tablewidth{0pt}
\tablecaption{Radio Spectral Parameters}
\tablehead{
\colhead{Cluster}& \colhead{Z\tablenotemark{a}}&
 \colhead{A(ergs cm$^{-2}$ sec$^{-1}$Hz$^{-1}$)\tablenotemark{b}} &
\colhead{Energy Spectral Index ($\alpha_{r}$)} & \colhead{Halo Radius (arc min)}
}
\startdata
	 A401\tablenotemark{c} & 0.0748 & 1.6$\times$10$^{-12}$	& 1.4 & 15	\nl
      A754\tablenotemark{d}	& 0.0528 & 2.5$\times$10$^{-12}$	&1.3 & 12 	\nl
      A1367\tablenotemark{e}	& 0.0213 & 5.5$\times$10$^{-11}$ & 1.5 & 8 \nl
	Coma\tablenotemark{f} & 0.0235 & 8.3$\times$10$^{-12}$ & 1.34 & 10 \nl
      A2256\tablenotemark{g}	& 0.0601 & 6.5$\times$10$^{-9}$	& 1.8\tablenotemark{i} & 5 \nl
      A2319\tablenotemark{h}	& 0.0529 & 2.0$\times$10$^{-11}$	& 1.4 & 5	\nl
\enddata
\tablenotetext{a}{Sarazin (1988)}
\tablenotetext{b}{S($\nu$)=A$\nu^{-\alpha_r}$}
\tablenotetext{c}{Roland et al. (1981): Flux of 0.080 Jy at 610 MHz}
\tablenotetext{d}{Andernach et al. (1988): 0.136 Jy at 2.7 GHz}
\tablenotetext{e}{Gavazzi (1978): 0.363 Jy at 610 MHz}
\tablenotetext{f}{Kim et al. (1990): 0.72 Jy at 1 GHz}
\tablenotetext{g}{Bridle and Fomalont (1976); Bridle et al. (1979): 0.1 Jy at 610 MHz.}
\tablenotetext{h}{Bridle et al. (1979) give a range of 1.2 $<$ $\alpha$ $<$ 1.8}
\tablenotetext{i}{Harris and Miley (1978): 1 Jy at 610 MHz}
\end{deluxetable}
\clearpage

\figcaption[]{In the upper panel, A401 GIS and HED data is shown with the best fit model. The model consists
of a Raymond and Smith thermal component with the maximum allowable
powerlaw added (90\% confidence). The model components are shown
separately and summed. In the lower panel $\chi^{2}$ vs. energy is shown.}\label{fig1}
\figcaption[]{Same as Figure 1 shown for A754.}\label{fig2}
\figcaption[]{Same as Figure 1 shown for A1367.}\label{fig3}
\figcaption[]{Same as Figure 1 shown for the Coma cluster.}\label{fig4}
\figcaption[]{Same as Figure 1 shown for A2256.}\label{fig5}
\figcaption[]{Same as Figure 1 shown for A2319.}\label{fig6}
\clearpage
\clearpage

\plotone{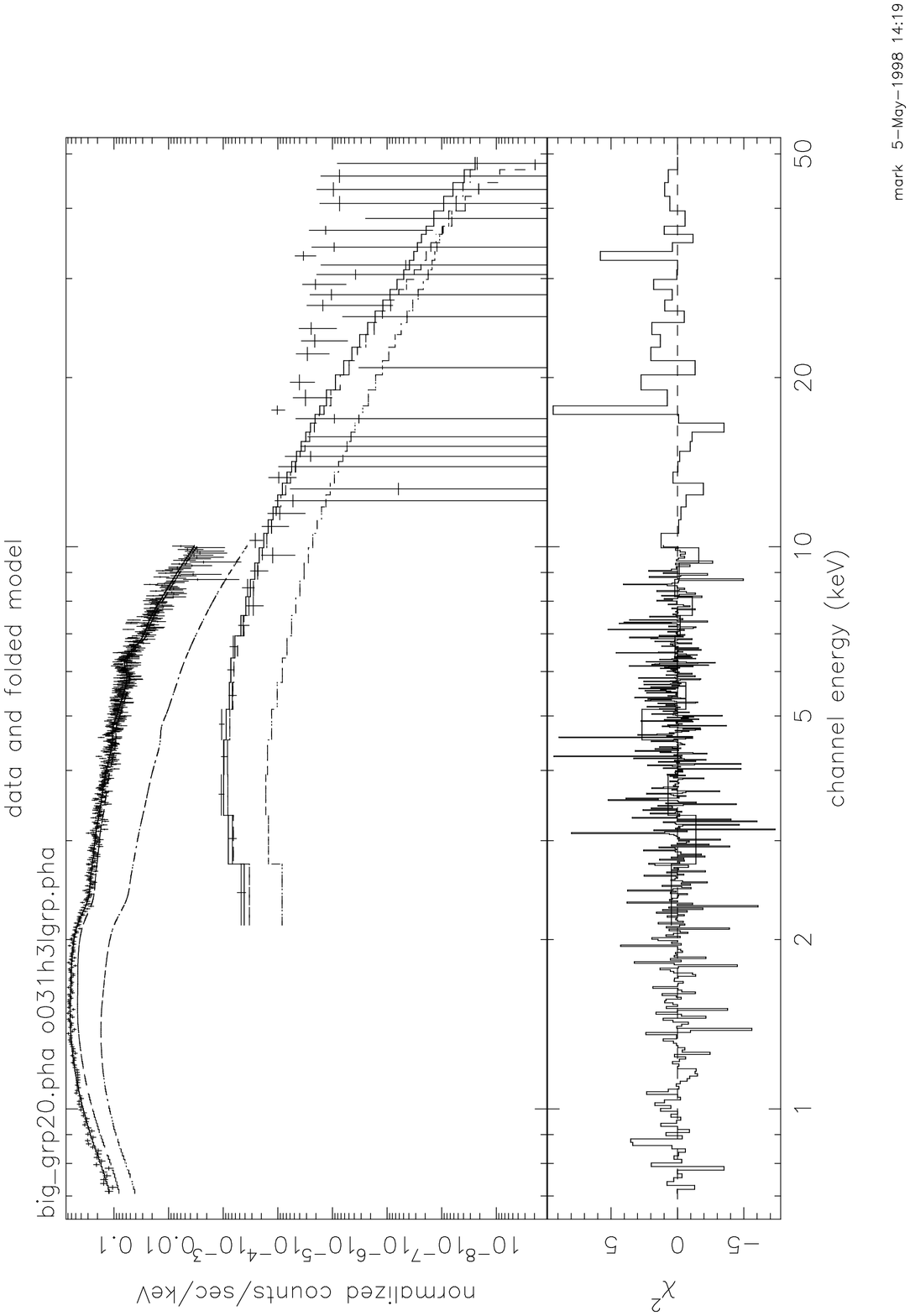}

\clearpage

\plotone{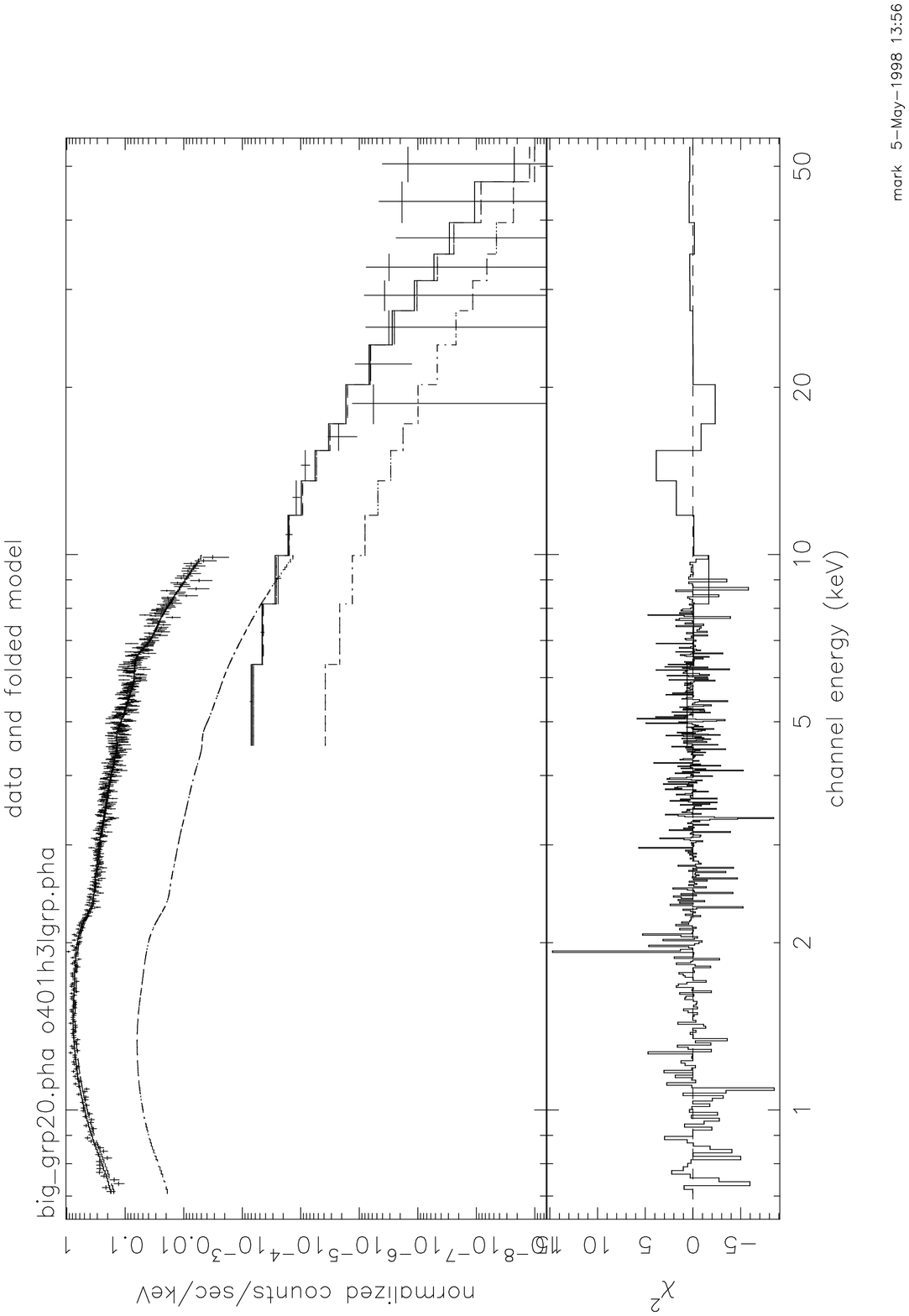}

\clearpage

\plotone{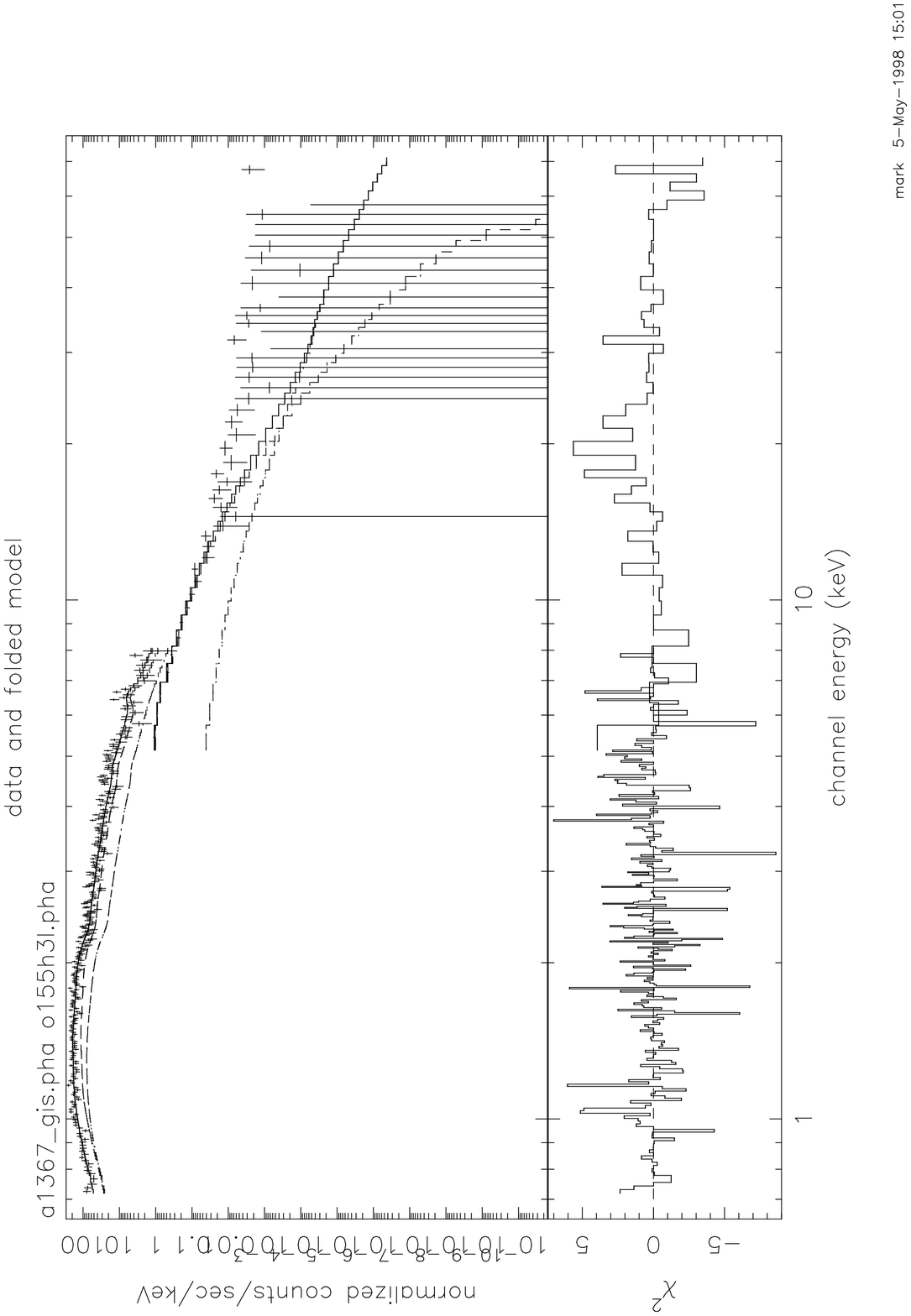}

\clearpage

\plotone{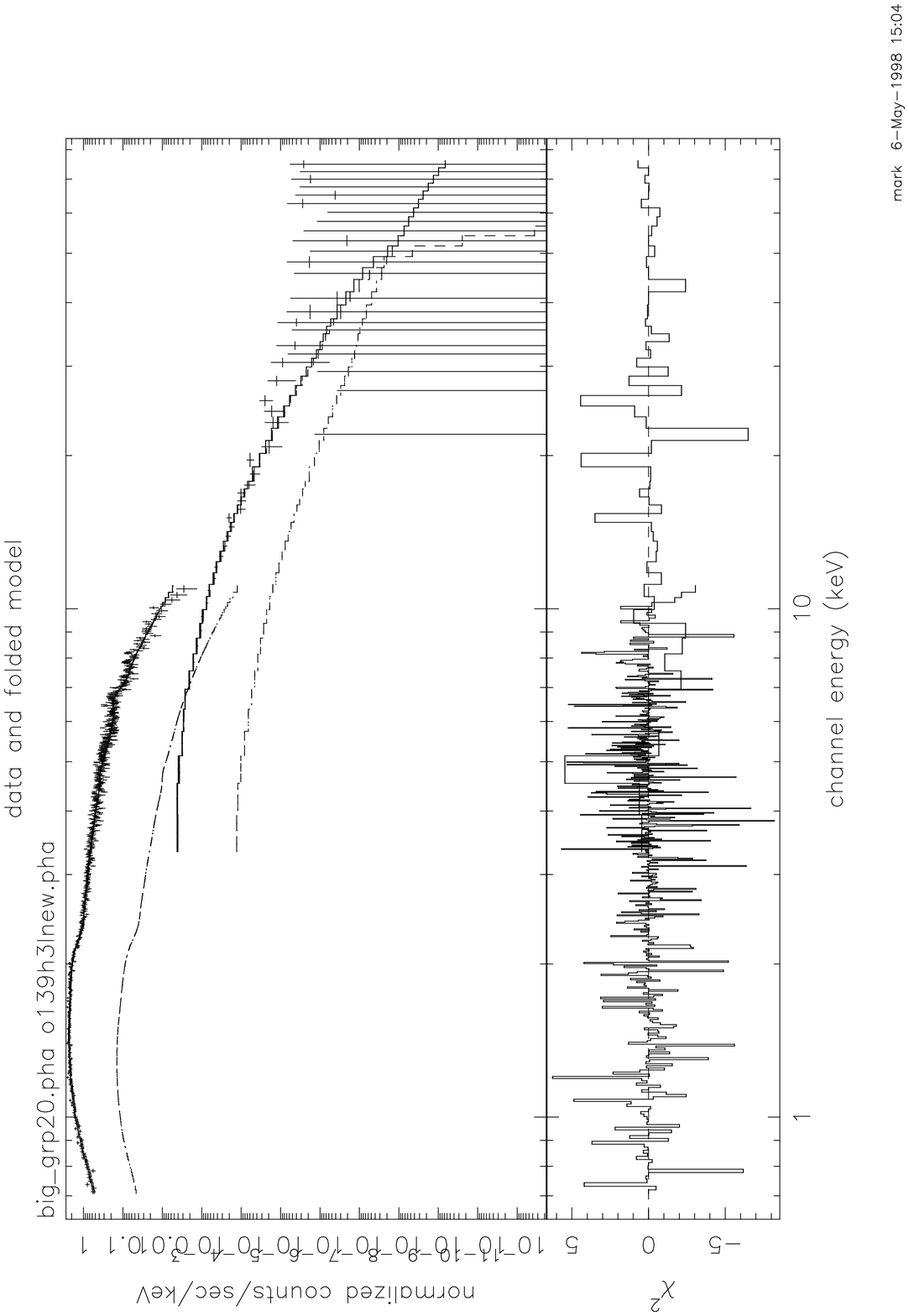}

\clearpage

\plotone{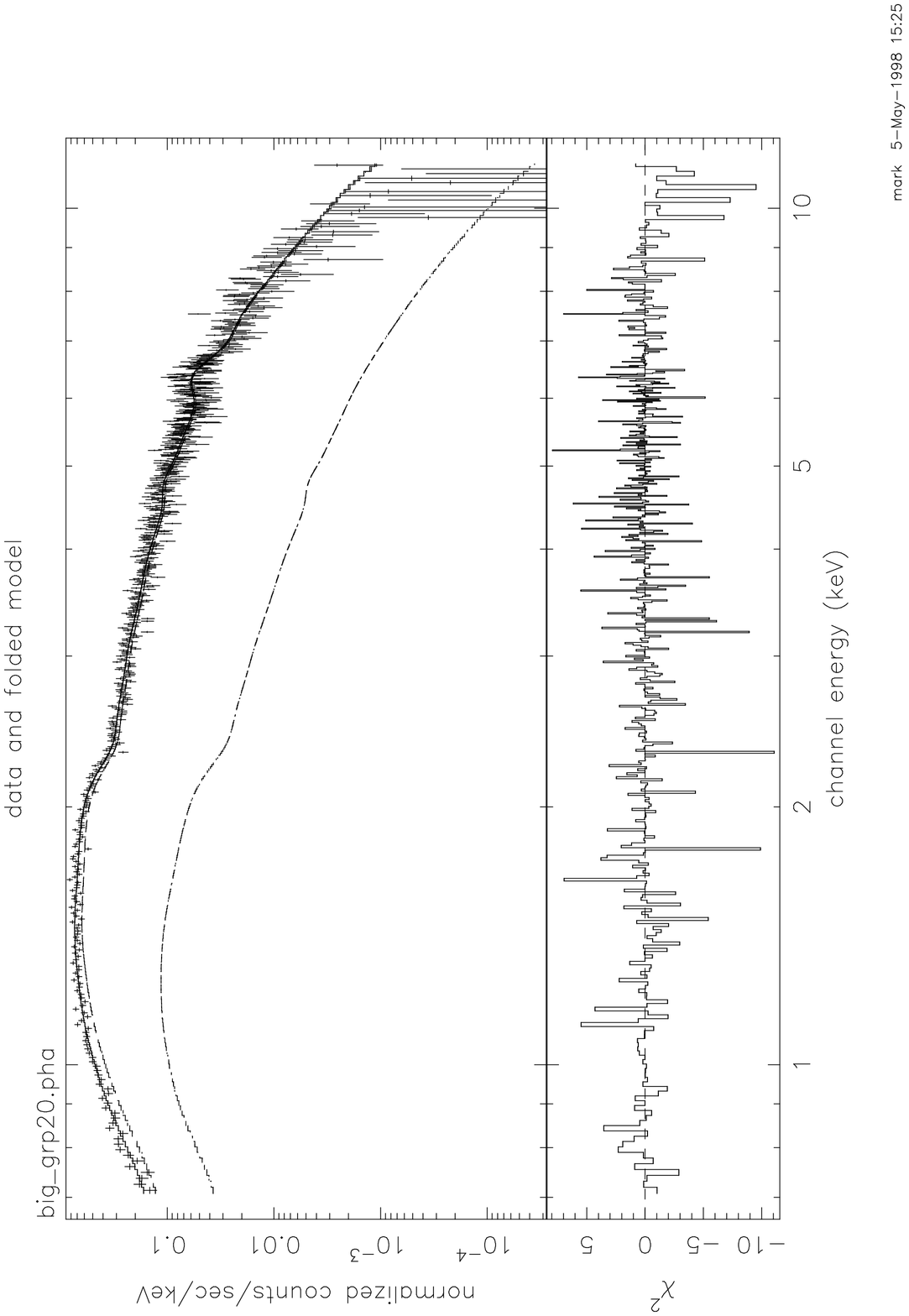}

\clearpage 

\plotone{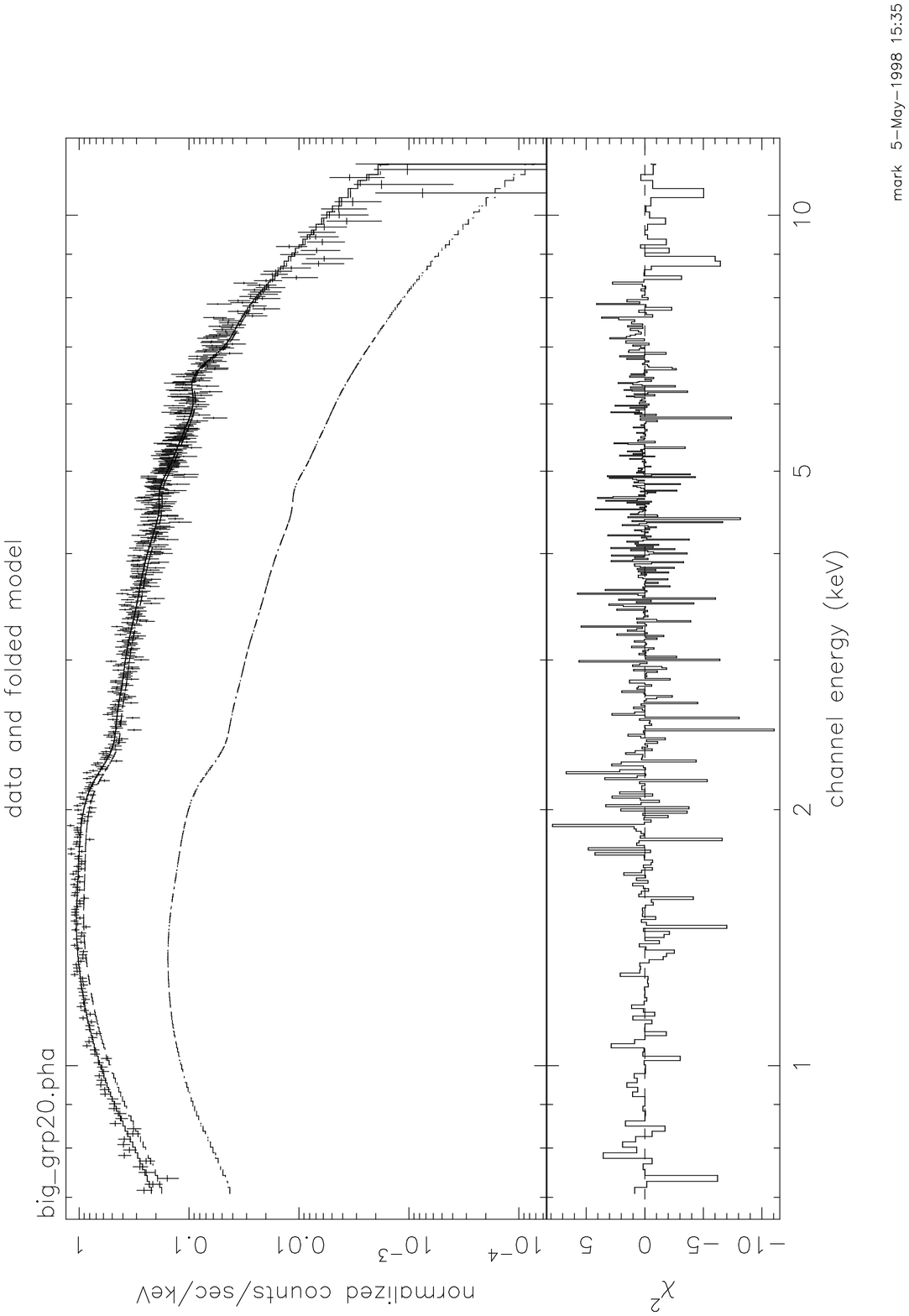}

\end{document}